\documentclass[12pt]{iopart}

\usepackage{graphicx}

\begin{document}

\title[]{Wavebreaking-associated transmitted emission of attosecond extreme-ultraviolet pulses from laser-driven overdense plasmas}

\author{Zi-Yu Chen$^{1,2}$, Mykyta Cherednychek$^{1}$ and Alexander Pukhov$^{1}$}
\address{$^{1}$Institut f\"ur Theoretische Physik I, Heinrich-Heine-Universit\"at D\"usseldorf, D\"usseldorf 40225, Germany}
\address{$^{2}$LSD, Institute of Fluid Physics, China Academy of Engineering Physics,
Mianyang 621999, China}
\ead{ziyu.chen@uni-duesseldorf.de or pukhov@tp1.uni-duesseldorf.de}
\vspace{10pt}
\begin{indented}
\item[]\today
\end{indented}

\begin{abstract}
We present a new mechanism of attosecond extreme-ultraviolet (XUV) pulses generation from a relativistic laser-driven overdense plasma surfaces in the wavebreaking regime. Through particle-in-cell simulations and analysis, we demonstrate that the observed ultrashort XUV emission for the parameters we considered is predominantly due to a strong plasma-density oscillation subsequent to wavebreaking. The coupling of the strong density variation and the transverse fields in the front surface layer gives rise to the transmitted emission with frequencies mainly around the local plasma frequency. This mechanism provides new insights into the scenarios of XUV generation from solid surfaces and the dynamics of laser-plasma interactions. 
\end{abstract}

%
%
%
\maketitle
%
%
\tableofcontents

\section{Introduction}
Laser-generated attosecond extreme-ultraviolet (XUV) pulses are of great interest as a powerful tool for a number of potential applications, including exploring novel ultrafast dynamics with unprecedented temporal resolution\cite{Krausz2009}, investigating nonlinear optics in the XUV region\cite{Heissler2012a}, and as a probe for laser-plasma interactions\cite{Borot2012,Dobosz2005}.

A promising approach to generate such XUV sources is high-order harmonics generation (HHG) from relativistic laser-irradiated overdense plasma surfaces\cite{Teubner2009}, which holds the potential and advantage to obtain XUV pulses with high brilliance. Several radiation mechanisms have been identified, such as coherent wake emission (CWE)\cite{Quere2006,Thaury2007,Thaury2010,Nomura2009}, relativistic oscillating mirror (ROM)\cite{Bulanov1994,Lichters1996,Baeva2006,Dromey2006,Pukhov2006,Dromey2007}, and coherent synchrotron emission (CSE)\cite{Brugge2010,Brugge2011,Dromey2012}. While CWE predominates at moderate laser intensities, i.e., the laser normalized vector potential $a \leq 1$, ROM and CSE are more efficient for highly relativistic laser intensities with $a\gg1$.  
So far, these mechanisms are well understood, and quite good agreement between the theoretical predictions and experiments has been achieved\cite{Quere2006,Thaury2007,Dromey2006,Dromey2007,Dromey2012}. 

In addition to HHG in the direction of laser reflection, transmitted XUV emission from the rear target surface have also been found, which is of interest for diagnostics to determine the maximum plasma density. Compared to the reflected HHG pulses, it also has the advantage of being used directly and not subject to additional energy loss due to further spectral filtering. 
Several mechanisms have been proposed\cite{Lichters1998,Gibbon1997,Hassner1997,Jarque1998,Teubner2004,Eidmann2005,Krushelnick2008,Chen2014,George2009,Dromey2013}, 
including line emission at twice of the plasma frequency due to inverse two-plasmon decay\cite{Lichters1998}, and emissions explained in the contexts of the CWE and CSE\cite{George2009,Dromey2013}. 
The analysis and demonstration of the above HHG mechanisms have greatly enhanced our understanding of the physics of laser solids interactions and XUV generation processes from solids surfaces\cite{Thaury2010}. 

In this paper, we report a new mechanism of ultrafast XUV pulse generation from laser-irradiated plasma surfaces in the wavebreaking regime. The attosecond XUV pulse is generated from the front layer of the plasma and then propagates through the foil target, with frequencies mainly around the local plasma frequency. Through simulations and analysis, we identify the underlying physics is predominately due to the strong plasma-density oscillation in the surface layer subsequent to wavebreaking, which we call \textit{wavebreaking-associated transmitted emission} (WTE). We also show that the emission is a general process for a wide range of laser and plasma parameters in the wavebreaking regime. Besides offering a new option to generate ultrafast XUV pulses, the radiation process identified here also provides important insights into the mechanism of XUV generation and the dynamics of laser-plasma interactions.  

\section{Simulation setup}
Both one-dimensional (1D) and 2D particle-in-cell (PIC) simulations are carried out using the Virtual Laser Plasma Lab (VLPL) code\cite{Pukhov1999}. We firstly present the 1D results to investigate the radiation mechanism in detail. Here, we demonstrate the basic idea mainly by considering the simplest configuration of normal laser incidence and step plasma density profile. As such, this mechanism can be pointed out most clearly since it is easily distinguished from the other mechanisms. For example, CWE cannot play a role with this geometry, because it requires oblique laser incidence and short density gradient\cite{Thaury2010}.
Besides, the emission cannot simply be attributed to ROM either, because ROM only occurs in the reflection direction\cite{Dromey2013}. 

The incident laser is linearly polarized in $z$-direction, with a Gaussian temporal
profile $a_{z}(t)=a_{L}\exp{(-t^{2}/\tau^{2})}$, where $a_{L}=eE_{L}/(m_{e}c\omega_{0})$ is
the normalized laser amplitude with $E_{L}$ and $\omega_{0}$
the laser field amplitude and the laser frequency respectively, $e$, $m_{e}$,
and $c$ are respectively the electron mass, the elementary charge and the speed
of light in vacuum, and $\tau=0.5T_0$ is the pulse duration with $T=2\pi/\omega_{0}$ the laser pulse duration. Here for simplicity we firstly consider this quasi-single-cycle pulse. The effect of multi-cycle laser pulses will be discussed in section 5.2. The fully ionized plasma, with a thickness of $d$=120 nm and an electron density of $n_{0}=100n_{c}$,
is initially located between $x=5\lambda_{0}$ and $x=5.15\lambda_{0}$,
where $\lambda_{0}$= 800 nm is the laser wavelength, $n_{c}=m_{e}\omega_{0}^{2}/4\pi e^{2}$
is the critical plasma density. The ions are assumed to be immobile
due to the short interaction time being considered. The cell size
is $\lambda_{0}/2000$ and each cell is filled with 100 macroparticles.

\section{Radiation features}
\begin{figure}\centering
\includegraphics[width=0.75\textwidth]{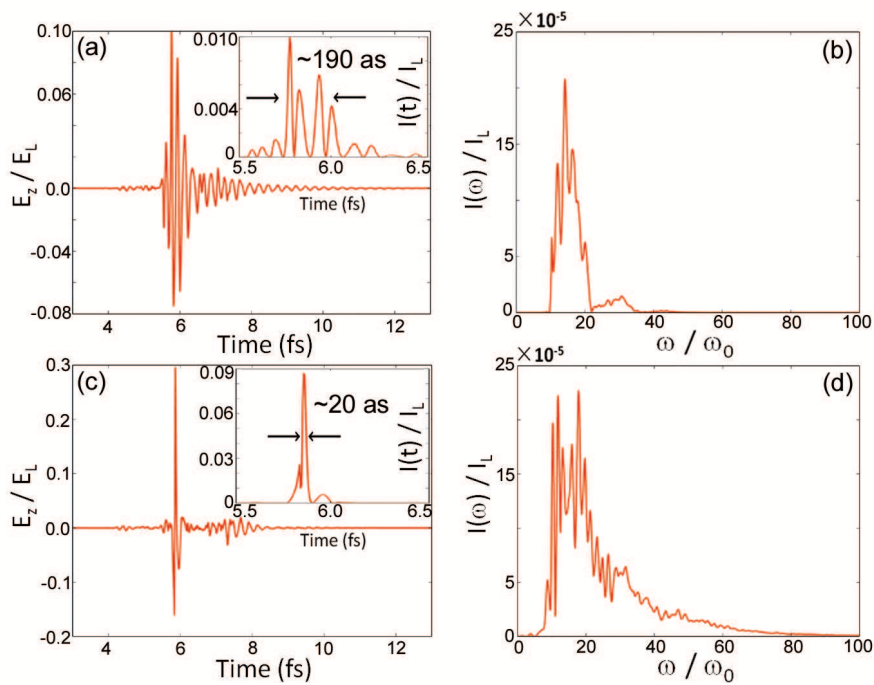}
\caption{\label{xuv_tf} (a), (c) Temporal profiles and (b),
(d) frequency spectra of the XUV pulses observed
at the rear side. Frames (a) and (b) are for $a_{L}=20$, and frames
(c) and (d) for $a_{L}=30$. The insets of (a) and (c) show the temporal
profiles plotted as intensity.}
\end{figure}

To show the essential signatures of the transmitted XUV pulses, we present
the results of two reference cases in Fig.~\ref{xuv_tf}, with frames (a) and (b) for a laser amplitude $a_{L}=20$, and frames (c) and (d) for $a_{L}=30$. From the temporal profiles of the electric field depicted in Fig.~\ref{xuv_tf}(a) and (c), one
can see intense few-cycle pulses have been generated. The emitted pulse has the same polarization with the laser pulse, i.e., only with electric field along $ z $-direction. This is different from CWE, which emits $y$-polarized HHG even for a $z$-polarized obliquely incidence laser\cite{Thaury2010}. The peak electric field is only about one order of magnitude smaller than that of the laser field. The transmitted XUV emission is strong. Its energy is about one-fifth of that carried by the reflected XUV harmonics (XUV frequency components from 10$\omega_0$ to 100$\omega_0$ are compared). For $a_{L}=30$, the XUV pulse is nearly single-cycle, reaching an extremely high peak field of $E_{z}/E_{L}=0.3$, corresponding to $E_{z}=3.5\times10^{13}$ V/m. The insets of Fig.~\ref{xuv_tf}(a) and (c) show the pulse intensity. The full-width at half-maximum (FWHM) of the pulses are approximately 190 attoseconds and 20 attoseconds for $a_{L}=20$ and $a_{L}=30$, respectively. 

Fig.~\ref{xuv_tf}(b) and (d) are the Fourier spectra corresponding to Fig.~\ref{xuv_tf}(a) and (c), respectively. The spectra display a low-frequency cutoff at the initial plasma frequency $\omega_{p0}=10\omega_{0}$. Thus, the pulse can be used directly as it is already filtered by the target. The pulse energy for $a_L=20$ as shown in Fig.~\ref{xuv_tf}(b) is mostly concentrated at the frequencies $\geq\omega_{p0}$. The spectrum for $a_L=30$ shown in Fig.~\ref{xuv_tf}(d) is broader and extends to higher frequencies. As such, the temporal pulse width in Fig.~\ref{xuv_tf}(c) is much shorter than that in Fig.~\ref{xuv_tf}(a). When we further increase the incident laser amplitude $a_{L}$ to above 40, the spectra show lower frequencies indicating laser light transmission, as a result of strong target compression by the light pressure.

\section{Radiation mechanism}

In this section we focus on revealing the underlying mechanism of the XUV emission. Here we select the case of $a_{L}=20$ for demonstration, since the basic features of radiation and the interaction dynamics are similar for both cases, as can be seen from the Supplemental movies SM1 and SM2\cite{SM}.

We start with discussing the possible origins of high-frequency emission based on the expression for total transverse current leading to the transverse radiation field. Considering a more general case of laser at oblique incidence with an angle of $\theta$, the expression for the radiation source, i.e., the total transverse current $\textbf{j}_{\tau}$, can be obtained as\cite{Thaury2010}
\begin{equation}\label{jtau}
\textbf{j}_{\tau}(x,t) = -\frac{e^2 n_e(x,t)}{m_e \cos \theta} \frac{\textbf{A}(x,t)}{\gamma(x,t)} -ec \tan\theta \Big[ Z n_i(x,t)-\frac{1}{\cos\theta} \frac{n_e(x,t)}{\gamma(x,t)} \Big] \hat{\textbf{e}}_y,
\end{equation}
where $\textbf{A}$ is the total vector potential, $Z$ is the ion charge number, $n_i$ is the ion density, and $\hat{\textbf{e}}_y$ is the unit vector along $y$-axis. 

The second term is always along the $y$-direction and only occurs for oblique incidence with $\theta \neq 0$. This is the source term responsible for the CWE mechanism\cite{Thaury2010}. It can emit radiation directly from the plasma oscillation $n_e$, without the need to couple with the transverse laser field. This is obviously not the case in our scheme where $\theta=0$ and the emission is in the $z$-direction. In addition to requiring oblique incidence, CWE is only possible in the presence of a density gradient, since its mechanism is basically the inverse process of resonance absorption.

Under the condition of laser field $E_z$ at normal incidence with $\theta=0$, the total transverse current is recast to be
\begin{equation}\label{jtau_normal}
\textbf{j}_{\tau}(x,t) = -\frac{e^2 }{m_e} \frac{n_e(x,t)}{\gamma(x,t)}\textbf{A}_z(x,t),
\end{equation}
In this case, high frequencies can be introduced by temporal modulation of the effective plasma density $n_e/\gamma$ and the vector potential $\textbf{A}_z$, and Doppler upshifting effect. The ROM mechanism is dominated by the Doppler upshifting effect, and thus can only occur in the reflected direction\cite{Dromey2013}. In the following, we show that our observed transmitted emission can be mainly attributed to strong density oscillations at wavebreaking level coupled to transverse electric fields in the laser-illuminated surface layer. 

\subsection{Region of emission}

\begin{figure}\centering
\includegraphics[width=0.9\textwidth]{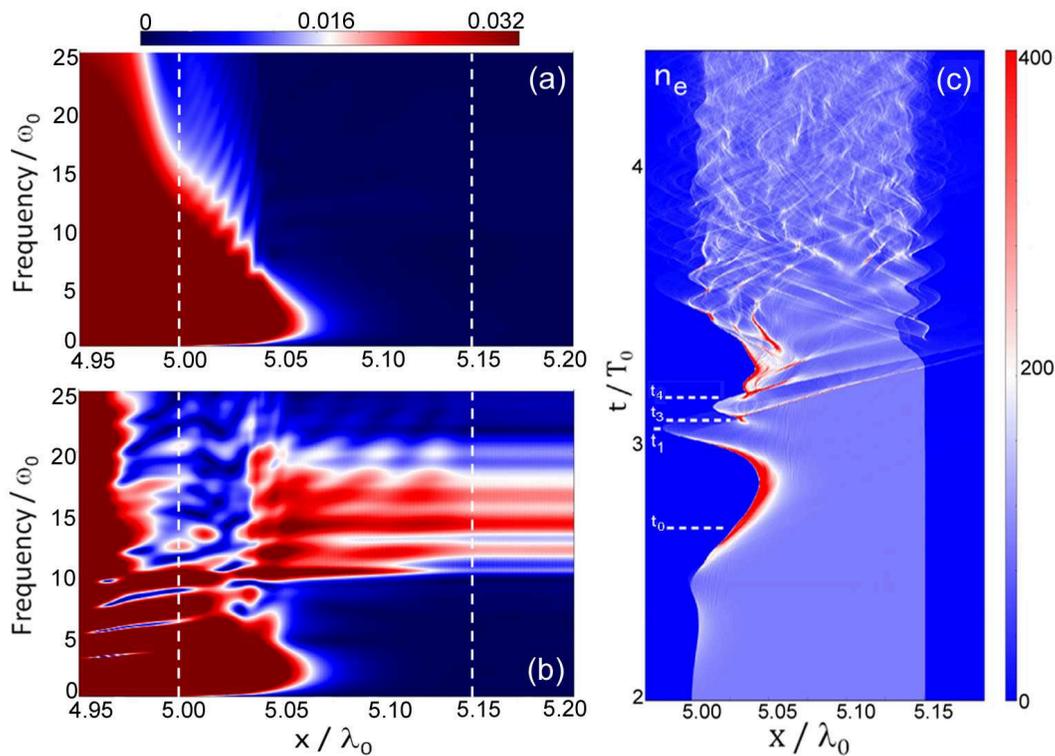}
\caption{\label{ez_f_x} (a)-(b) Spatially resolved spectrum of transverse
electric field $E_{z}$. At each spatial observation point $x$, Fourier transform is carried out with respect to the temporal waveform recorded. The time interval is from $t=0$ to $t=3.06T_0$ (marked as $t_1$ in frame (c)) for frame (a), and from  $t=3.06T_0$ to $t=8.0T_0$ for frame (b).  The vertical dashed white lines mark the initial plasma boundaries. (c) Spatial-temporal distribution of the electron density $n_{e}(x,t)$ in units of $n_{c}$. Time $t_3=3.10T_0$ is also marked, which indicates the onset of wavebreaking. Here, $a_L=20$.}
\end{figure}

We firstly demonstrate that the transmitted emission originates from the laser-illuminated front layer of the target. To see this, we plot the spatially resolved spectra in Figs.~\ref{ez_f_x}(a)-(b), which give the information about when and where the emission occurs. The procedure to obtain these spectra is as following. First, we record the temporal profile of the transverse electric field $E_{z}(x_0,t)$ at the spatial position $x_0$ over a period of time. Next, Fourier transformation is carried out with respect to this temporal profile $E_z(x_0,t)$ to obtain $E_z(x_0,\omega)$. We do this for each point of $x$ in the range between $x=4.95\lambda_0$ and $x=5.20\lambda_0$. Finally we map the spatial-spectra distribution of $E_z(x,\omega)$ as Figs.~\ref{ez_f_x}(a)-(b). The time interval for Fig.~\ref{ez_f_x}(a) is from $t=0$ to $t=3.06T_0$, and for Fig.~\ref{ez_f_x}(b) from $t=3.06T_0$ to $t=8.0T_0$. Here $T_0$ is the laser period. Time $t=3.06T_0$ is also marked as $t_1$ in Fig.~\ref{ez_f_x}(c), which shows the spatial-temporal distribution of the electron density $n_{e}(x,t)$. Time $t_3=3.10T_0$ marked in Fig.~\ref{ez_f_x}(c) indicates, as we will show later, the onset time of wavebreaking. This means that Fig.~\ref{ez_f_x}(a) and (b) are plotted respectively before and after the time when wavebreaking occurs. As can be seen, there is no transmitted emission before the wavbreaking occurs (see Fig.~\ref{ez_f_x}(a)), while after the onset of wavebreaking the transmitted emission is observed (see Fig.~\ref{ez_f_x}(b)). Note that the frequency of the transmitted emission in Fig.~\ref{ez_f_x}(b) accords with the XUV spectrum shown in Fig.~\ref{xuv_tf}(b). In addition to the timing of the transmitted emission, another important observation is that this XUV pulse observed at the target rear side originates from the front (laser-illuminated) layer of the target. The XUV pulse then propagates through the plasma slab.

Next, we show that strong plasma density oscillations occur in this front layer. From the spatial-temporal distribution of the electron density shown in Fig.~\ref{ez_f_x}(c), one can see that electrons at the front plasma surface are initially pushed forward by the laser light pressure and then bounce back in the first half laser cycle. The same process repeats in the second half laser cycle, but with some different features: in addition to follow the driving-laser-pulse shape, the plasma surface also exhibits higher frequency oscillations. As mentioned above, it is during this time the transmitted emission occurs.

\subsection{Onset of wavebreaking}

\begin{figure}\centering
\includegraphics[width=0.95\textwidth]{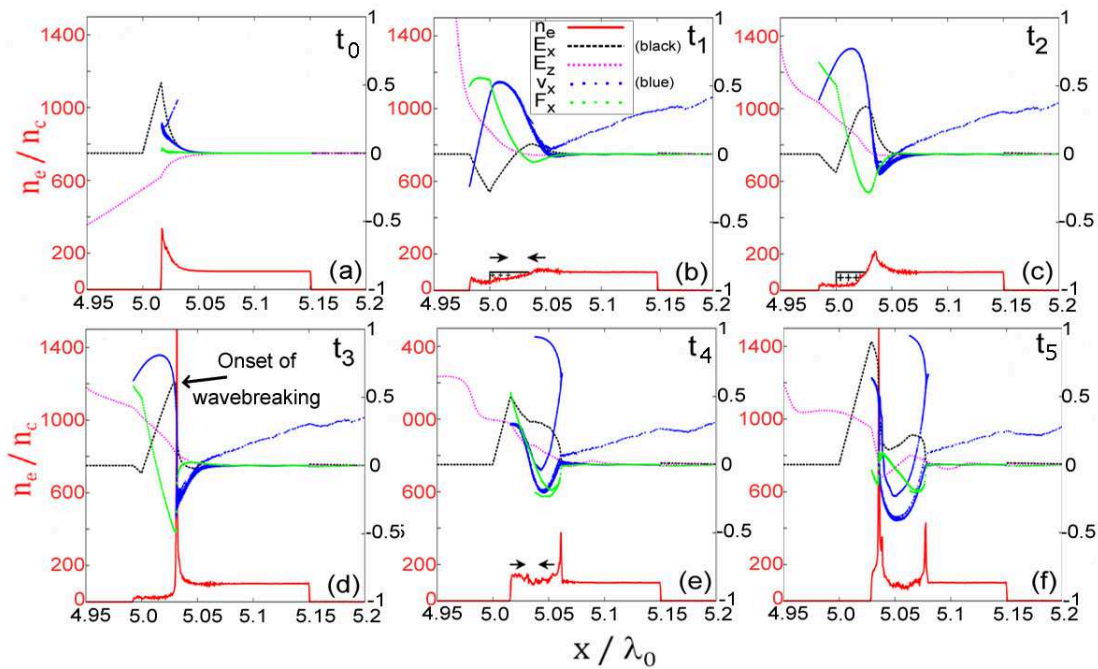}
\caption{\label{ne_ps} (a)-(f) Spatial profiles of $n_{e}$ (red), longitudinal electric field $E_{x}$ (black), transverse electric field $E_{z}$ (magenta), velocity in $x$-direction $v_{x}$ (blue), and $x$-component of the total force $F_{x}$ (green) at six reference times, $t_{0}=2.66T_{0}$, $t_{1}=3.06T_{0}$, $t_{2}=3.09T_{0}$, $t_{3}=3.10T_{0}$,
$t_{4}=3.16T_{0}=t_0+T_0/2$, and $t_{5}=3.18T_{0}$, where $T_0$ is the laser period. Here $a_L=20$. The very sharp high-density spike formed at $t_3$ and the onset of a multi-stream motion in the phase space are the signatures of wavebreaking.}
\end{figure}

To understand how the strong density oscillations arise, we present the plasma dynamics at six characteristic times in Fig.~\ref{ne_ps}: $t_0=2.66T_{0}$, $t_{1}=3.06T_{0}$, $t_{2}=3.09T_{0}$, $t_{3}=3.10T_{0}$, $t_{4}=3.16T_{0}=t_0+T_0/2$, and $t_{5}=3.18T_{0}$. 
In the first half laser cycle, due to the large laser ponderomotive force, surface electrons are pushed deep inside the plasma (see Fig.~\ref{ne_ps}(a)), creating a large electrostatic field. As the laser ponderomotive pressure passes its first maximum, the surface electrons are pulled back by the large electrostatic restoring force, gaining a large kinetic energy before exiting beyond the initial foil edge at $x=5\lambda_{0}$ (see Fig.~\ref{ne_ps}(b)). This group of electrons experiences a stronger inward acceleration in the second half of the laser cycle, when the ponderomotive pressure and electrostatic forces in the vacuum region are co-directed. When this group of electrons returns to the plasma edge, it meets background electrons that were initially deeper inside the surface layer and are now moving in the opposite direction. When the two groups of electrons cross
(see Fig.~\ref{ne_ps}(c)), a very sharp high-density spike forms at $t_3$ (see Fig.~\ref{ne_ps}(d)), also indicating the onset of a multi-stream motion in the phase space - \textit{signatures} of wavebreaking\cite{Dawson1959}. For initially cold plasma, the wavebreaking causes an extremely high spike in the local plasma density, although thermal pressure effects may limit the actual density increase\cite{Coffey1971,Schroeder2005}. 

\subsection{Strong density oscillation subsequent to wavebreaking}

The onset of wavebreaking is followed by a high level of plasma density oscillation. This can be seen from the spatial-temporal profile of the electron density shown in Fig.~\ref{ez_f_x}(c), and more clearly from the movie SM1 in the Supplemental Material\cite{SM}. The dynamic process of the density oscillation subsequent to wavebreaking can be understood as following. The electrons, which start to move in the negative direction at the beginning of the second half laser period, largely affect the following plasma oscillation. They represent the electron boundary after crossing over the electrons moving to the right. Afterwards, the laser ponderomotive force dominates and it reverses the boundary electrons (see Fig.~\ref{ne_ps}(e)). At the same time, the other electron bunches move further inside the plasma. This leads to the density profile largely different from that of half laser period ago when no wavebreaking occurs (see Fig.~\ref{ne_ps}(a)). In the case of without wavebreaking, $E_x$ decreases exponentially in the surface layer after reaching its maximum, while in the case of wavebreaking the electric field decays much slowly (see Fig.~\ref{ne_ps}(e)). As a result, the electrostatic force starts to dominate in the surface layer (see the total force) and consequently the electrons in this area start to move in the negative direction. At the same time, the electrons near the boundary still move in the positive direction, so a second sharp density spike is expected (see Fig.~\ref{ne_ps}(f)). This process repeats several times that several periods of density oscillation can be expected.


From the point of view of wavebreaking, a direct consequence of wavebreaking is a large number of formerly nonresonant main body electrons can rapidly exchange energy with the plasma wave and acquire momenta efficiently. In other words, wavebreaking implies a considerable fraction of the plasma electrons are trapped and start oscillating. This leads to a greatly increased plasma fluctuation level. The strong oscillation of many surface electrons can be seen from the rotating structure of the electron phase-space distribution at latter times, as shown by the blue dotted lines in Figs.~\ref{ne_ps}(f)-(e) and more clearly from the supplemental movie SM1\cite{SM}. 

\begin{figure}\centering
\includegraphics[width=0.9\textwidth]{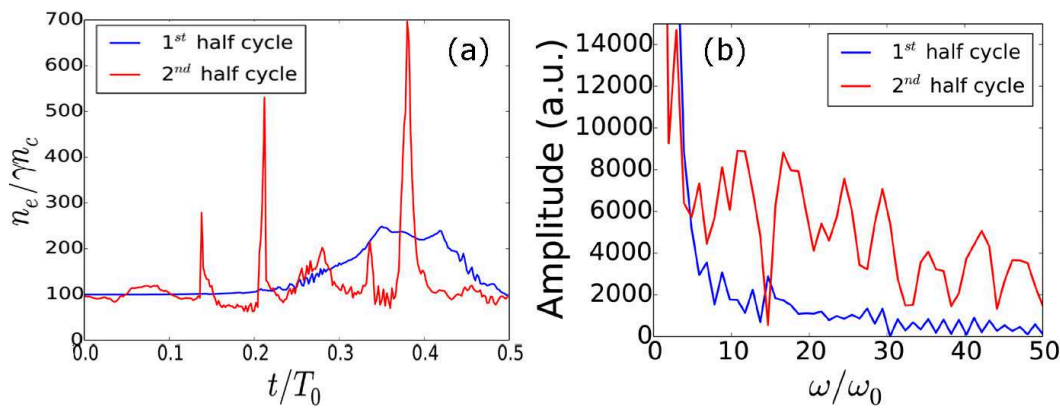}
\caption{\label{ne_tf} (a) The temporal profiles and (b) frequency spectra of the effective plasma density oscillation $n_e/\gamma$, recorded at a fixed position $x=5.05\lambda_0$ near the front surface. The blue lines correspond to a time interval between $t=2.5T_0$ and $t=3.0T_0$ (i.e., before the wavebreaking), while the red lines between $t=3.0T_0$ and $t=3.5T_0$ (i.e., after the wavebreaking).}
\end{figure}

The density variation is low and smooth before the wavebreaking, while high and fast after it.  For further demonstration, we present one example of the density variation recorded at a fixed position near the front surface $x=5.05\lambda_0$. Figures~\ref{ne_tf}(a) and (b) show the temporal profiles and the corresponding Fourier spectra, respectively. Here the (relativistically corrected) effective plasma density $n_e/\gamma$ is used and the Lorentz $\gamma$-factor is cell-averaged ($\gamma=<\gamma>_{cell}$). The blue lines in Fig.~\ref{ne_tf} correspond to a time interval between $t=2.5T_0$ and $t=3.0T_0$ (i.e., before the wavebreaking), while the red lines between $t=3.0T_0$ and $t=3.5T_0$ (i.e., after the wavebreaking). On the one hand, the amplitude of the density variation is much higher after the wavebreaking than before it (see the temporal profiles shown in Fig.~\ref{ne_tf}(a)). On the other hand, the frequency of the density oscillation is also higher after the wavebreaking, as shown in the frequency spectra of Fig.~\ref{ne_tf}(b).

\subsection{Emission of XUV pulses}

\subsubsection{Radiation source}
Although the plasma oscillation is longitudinal, it is coupled to the electromagnetic emission via the transverse velocity of electrons in the front surface layer. To understand this coupling, we consider the wave equation with source term (as shown in equation (\ref{jtau_normal})) in the 1D case. Assuming the laser potential $a_{z}^{laser}(x,t)$ in the skin layer is large, we can obtain the expression
for the small emitted wave $a_{z}^{e}$: 
\begin{equation}
\Big(\partial_{x}^{2}-\frac{1}{c^{2}}\partial_{t}^{2}\Big)a_{z}^{e}(x,t)\approx\frac{4\pi e^{2}}{m_e c^{2}}\left[\frac{n_{e}(x,t)}{\gamma(x,t)}-\frac{n_{0e}(x,t)}{\gamma_{0}(x,t)}\right]a_{z}^{laser}.\label{eq:wite}
\end{equation}
Here, $n_{0e}(x,t)$ and $\gamma_{0}(x,t)$ are the electron density and $\gamma-$factor in the skin layer respectively in the absence of the plasma oscillations. According to equation (\ref{eq:wite}), the emission is proportional to the laser amplitude and the amplitude of the electron plasma oscillation. 

The XUV amplitude increases with increasing the laser amplitude can be expected. To see the effect of density variation, we also did the same simulations by use of a lower intensity laser ($a_L=1$), or using a high intensity ($a_L=30$) but circularly polarized laser. In both cases, no wavebreaking and subsequent strong density oscillation can occur. As a result, no transmitted emission has been observed. Only in the wavebreaking regime, the electron density oscillation can reach such a high level that allows this kind of emission efficiently generated. It is also seen from the radiation source term in equation (\ref{eq:wite}) that the temporal variation of $n_e/\gamma$ is mainly responsible for the high-frequency XUV emission, since the laser field changes on a much longer time scale than the plasma oscillation assuming $n_e\gg n_c$. These demonstrate the strong density oscillation subsequent to wavebreaking indeed plays a dominant role in the emission process. This is why we call this radiation “wavebreaking-associated transmitted emission. 

\subsubsection{Simulated transverse current}
\begin{figure}\centering
\includegraphics[width=0.85\textwidth]{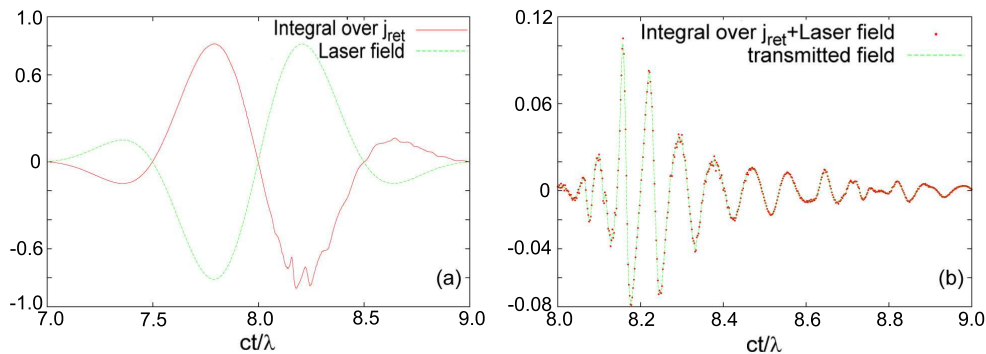}
\caption{\label{int_j} (a) The numerically calculated integral given by the first term in the RHS of equation (\ref{eq:intj}) (red) and the second term of the laser field $E_{laser}(t)$ (green). (b) The sum of the two colored plots given in frame (a) (red) and the transmitted field obtained directly from the PIC simulation results (green).}
\end{figure}

The density variation itself, though playing a dominate role in the emission process, does not fully determine the radiation properties. It is the retarded transverse current distribution that contains full information about the radiation. 
The transverse electric field measured from the rear side of the target is given by the general equation:
\begin{equation}\label{eq:intj0}
E_z(t,x)=-\frac{2\pi}{c}\int_{-\infty}^x j_z\left(t-\frac{x-x'}{c}, x'\right)dx' + E_{laser}(t,x).
\end{equation}
The first term in the right hand side (RHS) of equation (\ref{eq:intj0}) is the integral of the retarded transverse current distribution $j_z$ and the second term is the $z$-polarized laser field $E_{laser}(t,x)$ that propagates through vacuum.
To calculate the radiation at the end of the simulation box, we set $x=10\lambda$ and use dimensionless values. Thus we can drop the $x$-dependence of the fields and obtain:  
\begin{equation}\label{eq:intj}
E_z(t)=-2\pi\int_0^{x_{c}} j_z\left(t-x_{c}+x', x'\right)dx' + E_{laser}(t), 
\end{equation}
with $x_{c}=10$. Since the values of $j_z$ at each point of time and space can be obtained from the PIC simulation results, we can calculate the integral numerically as a function of time. The result is shown in Fig.~\ref{int_j}.

\begin{figure}\centering
\includegraphics[width=1\textwidth]{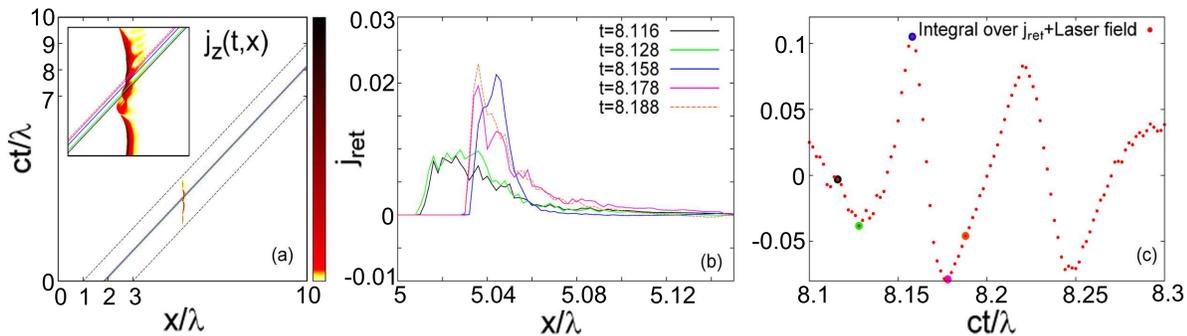}
\caption{\label{ret_j} (a) The current distribution $J_z(t,x)$ as a function of time and space. The colored oblique lines represent the paths along which $J_z(t,x)$ has to be integrated to calculate the transmitted field. (b) Spatial distribution of the retarded current density for five different times. Each line in frame (b) corresponds to the line of the same color in frame (a). (c) The transmitted fields calculated numerically using equation (\ref{eq:intj}). Each colored point corresponds to the current of the same color in frame (b).}
\end{figure}

During the first half laser cycle when no wavebreaking-associated plasma oscillation occurs, the value of the integral matches -$E_{laser}(t)$, i.e., the transmitted field is zero. During the second half laser cycle, the shape of the integral has some additional oscillations. Using equation (\ref{eq:intj}) we get the calculated transmitted field that matches the result obtained directly from the PIC simulations as expected (see Fig.~\ref{int_j}(b)).

The emission can also be seen directly from the evolution of $E_z$ obtained from the PIC simulations, as shown in Figs.~\ref{ne_ps}(a)-(f). $ E_z $ is an evanescent wave in the skin layer before the wavebreaking (see, e.g., Fig.~\ref{ne_ps}(a)). The distribution of $E_z$ changes drastically after the onset time of wavebreaking, e.g., as depicted in Fig.~\ref{ne_ps}(f), $E_z$ arises beyond the skin depth and propagates through the plasma foil. From the animated version of the field and plasma dynamics shown in movie SM1\cite{SM}), one can see more clearly that the transmitted emission occurs right after the onset of wavebreaking.

We can also analyze how the retarded current density oscillation evolves with time and how it corresponds to each value of the transmitted field. From the green line in Fig.~\ref{ret_j}(b), we can see that at time $t = 8.128 T_0$, the left boundary of the current profile reverses its moving direction and starts to move to the right. This leads to the transmitted field reaching a local maximum at the same time, as shown in Fig.~\ref{ret_j}(c). Similar process occurs at time $t = 8.158T_0$. The left boundary of the current profile reverses its moving direction again (see the blue line in Fig.~\ref{ret_j}(b)). Accordingly, the transmitted field reaches another local maximum(see Fig.~\ref{ret_j}(c)). Obviously, this behavior of the retarded current corresponds to the plasma-density oscillation we discussed above. The density oscillation leads to a fast variation of the total current and thus gives rise to radiation at higher frequencies that can propagate through the plasma.

\subsection{Theoretical analysis}
To enable the transmitted emission propagate through the plasma slab, the plasma should locally support higher frequency oscillations which then couple to the transverse electric fields. Here we give a simple model to show in principle it is possible for the local plasma frequency to be higher than the background plasma frequency due to density compression effect.
Let us consider the local plasma frequency at the laser-plasma interface. Let the plasma be
overdense with $\omega_{p}\gg\omega_{0}$. The laser is characterized
by its normalized vector potential $\textbf{a}(t)=e\textbf{A}(t)/m_{e}c^{2}$.
Electrons oscillate transversely in the laser field at the laser frequency
with the normalized momentum $\textbf{p}_{\perp}=\textbf{a}$. The
ponderomotive force is $\textbf{F}_{p}=-\nabla\gamma$\cite{Bauer1995,Macchi2013}, where the
gamma factor $\gamma=\sqrt{1+a^{2}}$. The ponderomotive force compresses
electron density within the skin depth. The force balance is then 
\begin{equation}
\textbf{F}=-n_{i}x-\frac{\partial\gamma}{\partial x}+\int_{x_{0}}^{x}n_{e}(x')dx'=0,\label{F}
\end{equation}
where $n_{i}$ and $n_{e}$ are respectively the ion and electron
density, $x_0$ indicates the position of the front of the electron skin layer (may be different from the initial position $x_{ini}=0$ due to the compression), and $x$ is a position inside the skin layer. Taking the spatial derivative
of equation (\ref{F}), we find the equilibrium electron density 
\begin{equation}
n_{e}(x)=n_{i}+\frac{\partial^{2}\gamma}{\partial x^{2}}.
\end{equation}
Thus, the equilibrium electron density within the skin layer is, as
expected, higher than the ion density due to the compression by the
laser ponderomotive force.

For simplicity, we consider laser field as quasi-static with respect
to the high plasma frequency. Thus, together with the electron-density
compression and the acquired relativistic $\gamma$-factor of electrons,
the local plasma frequency changes as well. It can be obtained from
the equation of motion $\mathrm{d}\textbf{p}/\mathrm{d}t=-\textbf{E}$
and the Maxwell's equation $\partial\textbf{E}/\partial t=n_{e}\textbf{p}/\gamma$\cite{Macchi2013},
where $\textbf{E}$ is the electric field, so that the local plasma
frequency is 
\begin{equation}\label{wpx1}
\omega_{p}(x)=\sqrt{\frac{n_{e}(x)}{\gamma(x)}}=\sqrt{\frac{\omega_{p0}^{2}+\partial^{2}\gamma/\partial x^{2}}{\gamma(x)}},
\end{equation}
where $\omega_{p0}=\sqrt{n_{i}}$. Because the local plasma frequency
has a strong spatial dispersion, plasma waves excited in this region
break easily. The change in the plasma frequency has two sources:
electron density compression and increase of the electron relativistic
$\gamma$-factor. These two effects tend to compensate each other,
but this compensation is incomplete. 

We assume an exponential decay of the laser field in the skin layer $a(x)=a_{s}e^{-\omega_p(x)(x-x_s)}$ in the vicinity of $x_s$, which is a position inside the skin layer. Considering within the skin layer the field amplitude can be small, we can write $\gamma\approx 1+a^{2}/2$.   
To proceed further, since $\omega_p(x)$ appears in the expression for $a(x)$, let us firstly find the zeroth-order approximation of $\omega_p$ near the position $x_s$. 
Now the $\gamma$-factor is constant and its derivative vanishes. 
Then equation (\ref{wpx1}) gives 
\begin{equation}
	\omega_p(x_s)\approx\frac{\omega_{p0}}{\sqrt{\gamma(x_s)}}\equiv\omega_{p1}
\end{equation}
To get the first-order correction, we can write $a(x)\approx a_{s}e^{-\omega_{p1}(x-x_s)}$ and set $x=x_s$ after calculating the second derivative of $\gamma$.
In this case we obtain 
\begin{equation}
	\omega_{p2}=\sqrt{\frac{\omega_{p0}^2+2a_s^2\omega_{p1}^2}{\gamma(x_s)}}.
\end{equation}
For the next higher-order approximations, we can insert $\omega_{p2}$ instead of $\omega_{p1}$ in the expression for $a(x)$ and so on in the same way. 
Consequently we obtain the sequence  
\begin{equation}
	\omega_{p~n+1}=\sqrt{\frac{\omega_{p0}^2+2a_s^2\omega_{pn}^2}{\gamma(x_s)}},\quad n=1,2,...
\end{equation}
The limit of this sequence is given by  
\begin{equation}
	\lim_{n\rightarrow\infty}\omega_{pn}=\frac{\omega_{p0}}{\sqrt{\gamma(x_s)-2a_s^2}}\approx\frac{\omega_{p0}}{\sqrt{1-\frac{3}{2}a_s^2}}.
\end{equation}
Therefore we arrive at the approximate local plasma frequency at the position $x_s$ under the assumption of small field amplitudes within the skin layer:
\begin{equation}
\omega_p(x_s)\approx \frac{\omega_{p0}}{\sqrt{1-\frac{3}{2}a_s^2}}.
\end{equation}
This expression is larger than the background plasma frequency mainly due to density compression. Thus, the local plasma oscillations can excite electromagnetic waves at frequencies above the background plasma frequency. These waves can propagate through the plasma slab and exit from the rear side of the target. When the laser intensity is too high, plasma oscillations can have a fundamental frequency blow that of the background. However, if the oscillations are nonlinear, their harmonics can propagate through.

\section{Discussions}

\subsection{Parametric study}

\begin{figure}\centering
\includegraphics[width=1\textwidth]{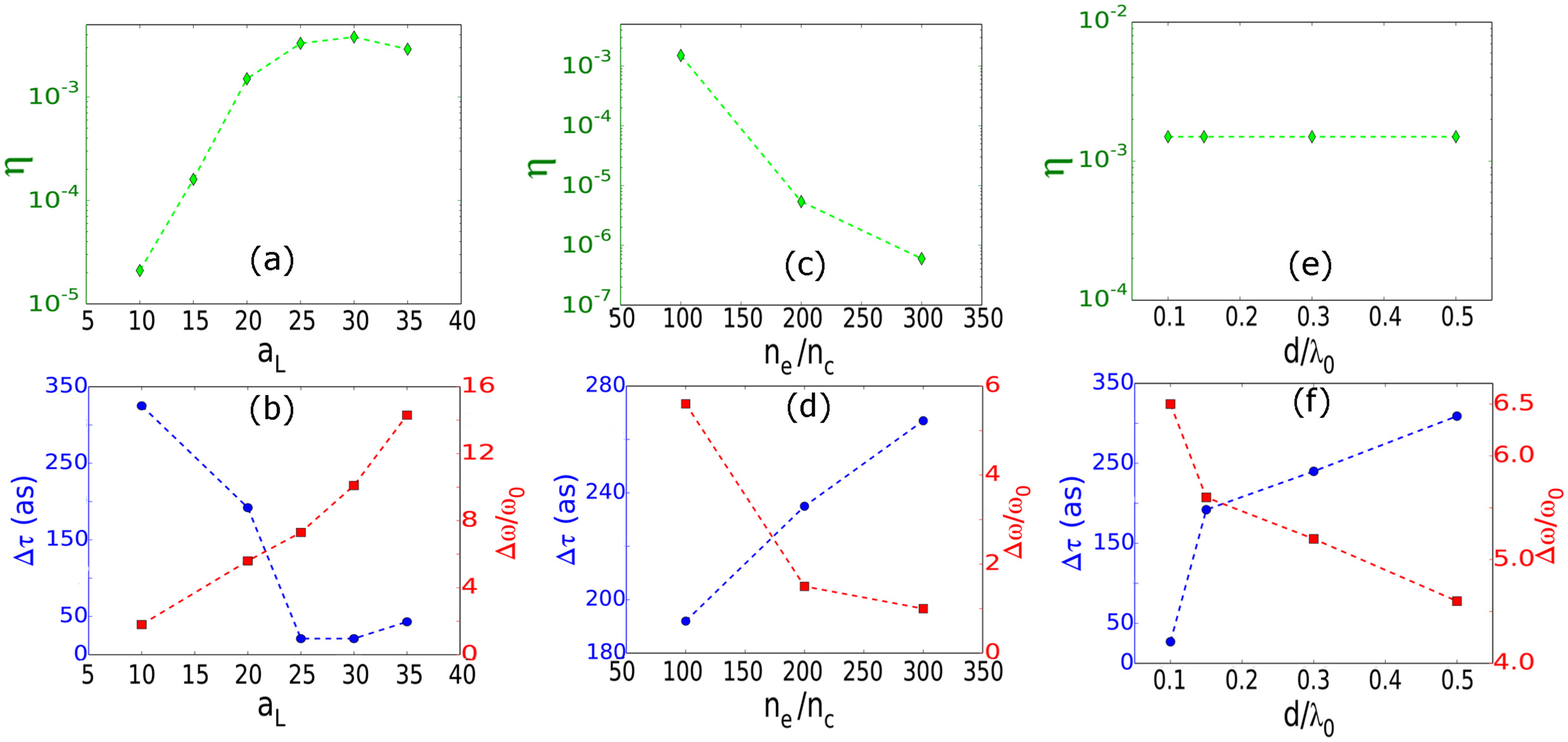}
\caption{\label{param} Influences of laser normalized amplitude $a_L$ (a-b), initial plasma density $n_e$ (c-d), and foil thickness $d$ (e-f) on the XUV energy conversion efficiency $\eta$, XUV pulse FWHM width $\Delta\tau$, and XUV pulse FWHM bandwidth $\Delta \omega$. In these simulations, when one parameter is varied, the other parameters are the same with that presented in section 2. The laser amplitude is $a_L=20$ in (c)-(f).}
\end{figure}

Here we present a systematic study to show how the parameters such as laser intensity, initial plasma density and target thickness influence the generated XUV pulses. Figure \ref{param}(a) shows the energy conversion efficiency $\eta$ grows with the normalized laser amplitude $a_L$, reaching about $4\times10^{-3}$ when $a_L=30$. Here, $\eta=\int E^2(\xi_{tr}) d \xi_{tr}/\int E^2(\xi_0) d \xi_0$, with $\xi=t-x/c$, $\xi_{tr}$ and $\xi_0$ denoting the transmitted and incident pulses, respectively. When $a_L$ further increases, $\eta$ decreases and the foil becomes more transparent to the laser pulse. Figure \ref{param}(b) suggests that higher laser intensity also favors the generation of XUV pulse with shorter FWHM duration $\Delta\tau$, and correspondingly broader FWHM bandwidth $\Delta \omega$. Figure \ref{param}(c)-(d) show the influence of the initial plasma density $n_e$. While $\Delta\tau$ increases and $\Delta \omega$ decreases with the increase of $n_e$, the conversion efficiency $\eta$ drops rapidly. This is because wavebreaking and strong density oscillations are more difficult to drive for higher density plasmas. The laser intensity should also increase with the plasma density in order to keep the laser plasma dynamics the same, as indicated by the dimensionless similarity parameter $S=n_e/a_L n_c$ from the similarity theory in the ultrarelativistic regime $a_L^2 \gg 1$\cite{Gordienko2005}. For the foil thickness $d$, it does not affect the conversion efficiency, as shown in Fig. \ref{param}(e). This is understandable because the generation process occurs at the front layer of the target. However, $\Delta\tau$ increases with $d$ while $\Delta \omega$ decreases (see Fig. \ref{param}(f)). This can be attributed to a result of dispersion when the pulses propagate through the plasmas.

\subsection{Effect of laser pulse duration}

\begin{figure}\centering
\includegraphics[width=0.75\textwidth]{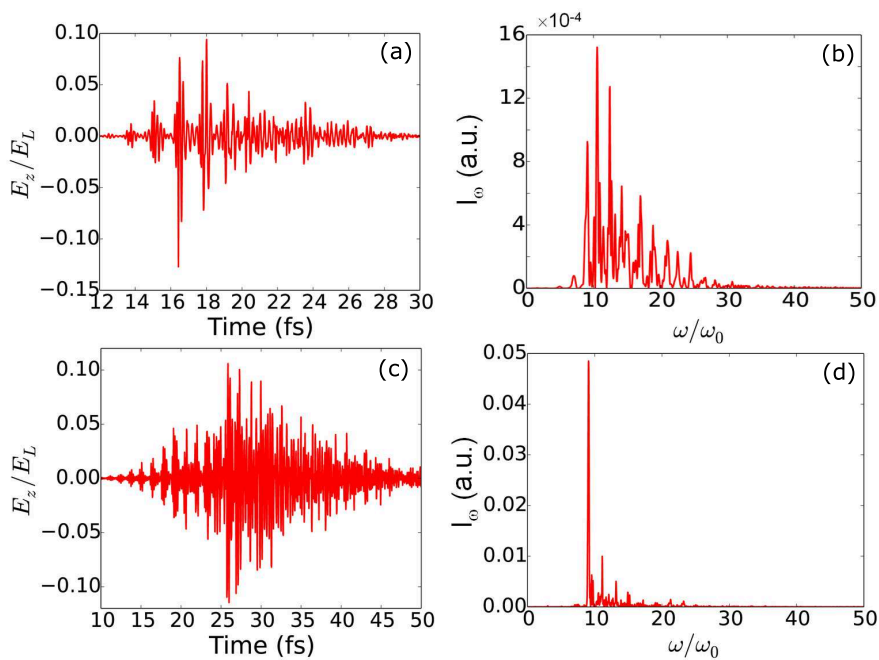}
\caption{\label{mul-cycle} Temporal profile and the corresponding Fourier spectrum of the transmitted XUV emission for the case of using multi-cycle laser pulses. The laser pulse durations are 10 fs for frames (a)-(b) and 30 fs for (c)-(d). The other parameters are the same with that in Fig.~\ref{xuv_tf}(a).}
\end{figure}

Next we examine the WTE generation by considering two of the most concerns in real experimental cases, i.e., the effect of using multi-cycle laser pulses and presence of a finite plasma density gradient, to demonstrate the robustness of this mechanism. 

Figure~\ref{mul-cycle}(a)-(b) show respectively the temporal profile and the Fourier spectrum of the WTE using a laser pulse with duration of 10 fs. The other parameters are the same with the case of $a_L=20$ in the above simulations. It is seen that a train of attosecond XUV pulses have been generated. The frequency spectrum contains finer structures than that in Fig.~\ref{xuv_tf}(b), as a result of the interference between different pulses in the attosecond pulse train. The effect of interference is more evident with a longer duration pulse of 30 fs, as shown in Figs.~\ref{mul-cycle}(c)-(d). Frequency components below $10\omega_0$ can be attributed to a lowered plasma frequency of the relativistically heated foil. Using techniques such as polarization gating, an isolated attosecond XUV pulse may be obtained. Nevertheless, these results shows the WTE mechanism also works by use of multi-cycle laser pulses. 

\subsection{Effect of plasma density gradient}

\begin{figure}\centering
\includegraphics[width=0.75\textwidth]{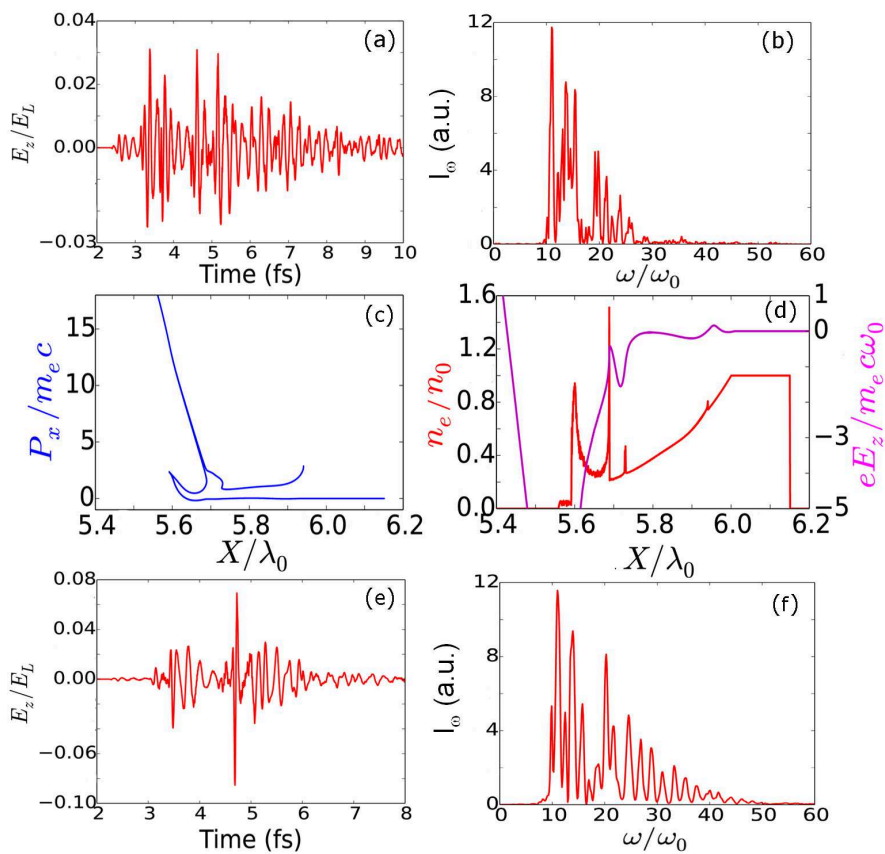}
\caption{\label{pr_T} (a) Temporal profiles and (b) Fourier spectra of the transmitted XUV emission for an exponential plasma density gradient with scale length $L =\lambda_0/5$. (c) Phase space distribution ($x,P_x$) and (d) profiles of the electron density $n_e$ and transverse field $E_z$ at time $t=3.4T_0$ for the case of $L =\lambda_0/5$. (e) Temporal profiles and (f) Fourier spectra of the transmitted XUV for the case of $L =\lambda_0/10$. The other parameters are the same with that in Fig.~\ref{xuv_tf}(a).}
\end{figure}

Since the WTE process relies on the strong plasma-density oscillation subsequent to wavebreaking, a finite density ramp in the front of the plasma surface will affect the threshold of wavebreaking and thus the temporal and spectral structures of the WTE. From the cold nonrelativistic wavebreaking field $E_{WB} = m_e c \omega_p/e$, which is dependent on the plasma frequency and thus the plasma density, we see that a density ramp can lower the wavebreaking threshold. Consequently, the presence of a pre-plasma allows the WTE to occur more easily. This can be seen by considering a longer plasma gradient length. Figure~\ref{pr_T}(a) shows the temporal profile of the WTE for the case of an exponential plasma density gradient with scale length $L =\lambda_0/5$. The emitted pulse lasts a longer time of several femtoseconds, compared with the case of without pre-plasma in Fig.~\ref{xuv_tf}(a). This is due to an earlier wavebreaking and the subsequent strong density oscillation when the laser pulses interact with the density ramp in the front of the target, as shown in Figs.~\ref{pr_T}(c)-(d). We can see the multi-stream motion of electrons from the phase space distribution ($x,P_x$), indicating the wavebreaking has occurred (see Fig.~\ref{pr_T}(c)). At the same time, the profile of the transverse field $E_z$ shows the waveform with higher frequencies (see Fig.~\ref{pr_T}(d)). Compared to the case of $L =\lambda_0/5$, the duration of the emitted pulse is shorter for the case of a shorter scale length of $L =\lambda_0/10$, as shown in Fig.~\ref{pr_T}(e). Correspondingly, the bandwidth of the spectrum in Fig.~\ref{pr_T}(f) is broader than that in Fig.~\ref{pr_T}(b).

\begin{figure}\centering
\includegraphics[width=0.75\textwidth
]{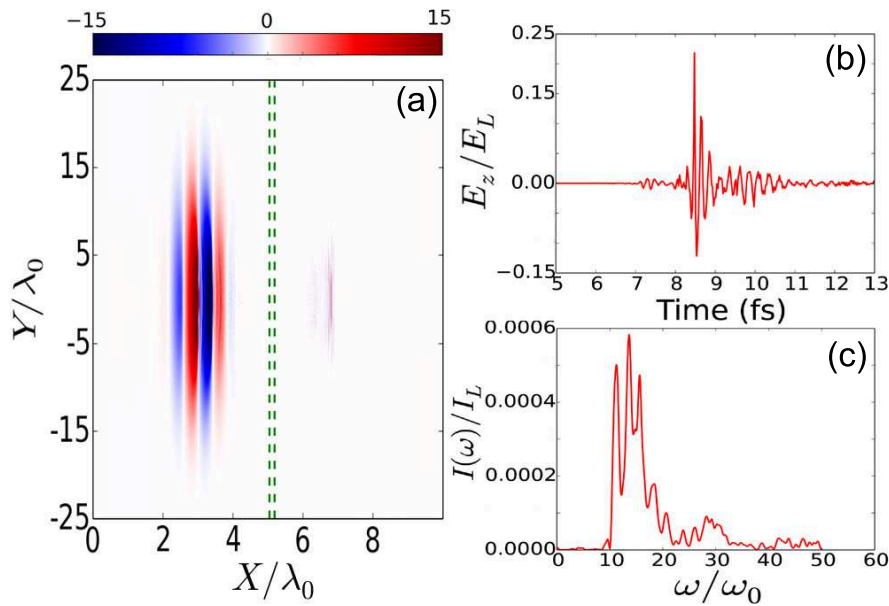}
\caption{\label{2d} 2D simulation results. (a) A snapshot of the transverse electric field $a_z$ distribution at simulation time $t=5T_0$. The green dashed lines mark the initial plasma boundaries. (b) Temporal waveform and (c) frequency spectrum of the transmitted emission recorded at (x=6$\lambda_0$, y=0).}
\end{figure}

\subsection{Multi-dimensional effects}
The results obtained so far are based on the 1D case. We also performed 2D simulations to check whether this mechanism works in multidimensional cases. Here we only intend to compare with the 1D results of the case shown in Fig.~\ref{xuv_tf}(a)-(b) to validate the basic physics. In the simulation, a very small grid step of $\lambda_0$/1000 is used in the $x$ direction in order to resolve the wavebreaking related process. Each cell is filled with 8 macroparticles. The other laser and plasma parameters are the same with those used in Fig.~\ref{xuv_tf}(a)-(b) in the 1D simulations, except that the laser pulse has a Gaussian transverse profile with a focal spot size of $10\lambda_0$. Figure~\ref{2d}(a) shows a snapshot of the transverse electric field distribution at time $t=5T_0$. We see that an ultrashort pulse is generated at the rear side of the target. Figures~\ref{2d}(b)-(c) show the temporal profiles and the corresponding frequency spectra of the transmitted emission observed at the position (x=6$\lambda_0$, y=0) at the rear target side, respectively. The signatures of both the temporal waveform and frequency spectrum are in good agreement with the 1D simulation results. An animated demonstration of the laser-plasma dynamics similar with the 1D case can be seen in movie SM3 in the Supplemental Material\cite{SM}. The density spike indicating the onset of wavebreaking at $t=3.11T_0$ and the subsequent transmitted emission at later times can be clearly seen from the movie. As for the concern of transverse instabilities, we note that experiments of femtosecond-picosecond laser interaction with nanometer-micrometer thin target are routinely available nowadays. A number of experiments have been carried out using similar parameters with ours. For example, experiments of a much longer pulse laser of 500 fs duration ($a_L\approx 20$) interacting with thin targets of thickness 125-200 nm have successfully demonstrated the transmitted emission due to the CSE mechanism\cite{Dromey2012}. These results indicate the transverse instabilities are not fatal with the parameters we considered here.

\section{Conclusions}
In conclusion, a new regime of attosecond XUV pulses generation from overdense plasma surfaces, namely, wavebreaking-associated transmitted emission (WTE), has been demonstrated. The emission originates from the plasma front surface and propagates through the target, with frequencies mainly around the local plasma frequencies. The underlying physics can be attributed to the coupling of the transverse fields in the skin layer and the strong plasma-density oscillation subsequent to wavebreaking. Thus the emission is evident only in the wavebreaking regime. This novel scenario of ultrafast XUV pulse emission from overdense plasmas provides new insights into the dynamics of laser-plasma interactions and the physics of radiation process. It may also offer an alternative option to generate relativistically intense ultrashort XUV pulses that may find extensive applications.

\section*{Acknowledgments}
We would like to thank Dr. John Farmer for helpful discussions. 
Z. Y. C. acknowledges financial support from the China Scholarship Council (201404890001). 
This work was supported by the Deutsche Forschungsgemeinschaft SFB TR 18, EU FP7 project EUCARD-2, and the Science and Technology Fund of the National Key Laboratory of Shock Wave and Detonation Physics (China) with project Nos. 077110 and 077160.

\end{document}